# A 137.5 TOPS/W SRAM Compute-in-Memory Macro with 9-b Memory Cell-Embedded ADCs and Signal Margin Enhancement Techniques for AI Edge Applications

Xiaomeng Wang[1,2], Fengshi Tian[1,2], Xizi Chen[4], Jiakun Zheng[1,2], Xuejiao Liu[2], Fengbin Tu[1,2], Jie Yang[3], Mohamad Sawan[3], Kwang-Ting (Tim) Cheng[1,2], Chi-Ying Tsui[1,2]

[1]The Hong Kong University of Science and Technology, Hong Kong, [2]The AI Chip Center for Emerging Smart Systems, Hong Kong, [3]Westlake University, Hangzhou, China, [4]Huazhong Agricultural University, Wuhan, China

Recent compute-in-memory (CIM) designs [1-6] have highlighted the potential for energy-efficient analog MAC computation using SRAM. Modern CNN applications require high-precision input (IN), weight (W) and output (OUT) to ensure inference accuracy [1]. However, the costly analog-to-digital converters (ADCs) hinder high energy efficiency in CIM designs. Moreover, as the number of MAC accumulations increases, signal margin degradation occurs, leading to decreased readout accuracy [7]. The current high-precision CIM macro designs are unable to achieve an optimal balance between energy efficiency and accuracy. Fig. 1 shows a comparison of our proposed design with the state-of-the-art CIM macro designs in terms of parallelism, accuracy, and energy efficiency. In previous works [2-4] and [6], in-memory MAC operations are performed on 2-bit ACT and 1-bit weight, with a limited number of accumulations to ensure signal margin when using a low-precision ADC. However, this approach needs multiple cycles of MAC-ADC and digital shift-and-add operations to achieve high-precision output, which reduces the parallelism of the CIM core and degrades overall energy efficiency. In [5], parallel multiplication of high-precision activations and weights is achieved by using multiple metal-oxide-metal (MOM) capacitor ladders in the memory sub-arrays. Accumulation is performed through charge-averaging-based summation among the sub-arrays before ADC, which reduces the amount of digital accumulation and amortizes the energy overhead of an 8-bit SAR-ADC. However, this method compromises computation accuracy due to degraded signal margin resulting from charge averaging.

In this paper, we propose a high-precision SRAM-based CIM macro that can perform 4×4-bit MAC operations and yield 9-bit signed output. The inherent discharge branches of SRAM cells are utilized to apply time-modulated MAC and 9-bit ADC readout operations on two bit-line capacitors. The same principle is used for both MAC and A-to-D conversion ensuring high linearity and thus supporting large number of analog MAC accumulations. The memory cell-embedded ADC eliminates the use of separate ADCs and enhances energy and area efficiency. Additionally, two signal margin enhancement techniques, namely the MAC-folding and boosted-clipping schemes, are proposed to further improve the CIM computation accuracy.

Fig. 2 illustrates the proposed CIM architecture and workflow of the MAC and readout processes. The design consists of 4 analog CIM cores, each with 4Kb 9-T SRAM cells. A CIM core comprises 16 column-wise dot-product CIM engines that store 64 weight data, each 4 bits wide, along with a sign control logic. The 9-T cells are partitioned into MAC and readout cells, as both processes utilize the embedded cells to perform a discharge-based operation on two matched MOM capacitors at read bit-lines (RBLs). During the AD process, a sense amplifier (SA) connected to the RBL and RBLB pair compares two bit-line voltages. The digital-to-time converter (DTC) and pulse path configuration circuits serve all the CIM engines during the MAC and AD operations. The 9-T SRAM cell comprises a 6-T cell that stores 1-bit weight data and three transistors functioning as a discharging branch. To enhance energy efficiency and slew rate, the active-low input pulse is applied to the source node instead of the gate of $M_0$ due to its small parasitic capacitance. A long-channel transistor $M_0$ is utilized to mitigate channel-length modulation and mismatch effects. Fig. 2 also shows the definition of signal margin (SM) as the difference between the MAC step size and the MAC result variance ($\mu_0$-2σ). The MAC step size ($\mu_0$) is determined by the ratio of the MAC voltage headroom and the dynamic range of the MAC results ($VPP_{MAC}/\sum MAC$). To increase the signal margin and achieve higher accuracy, a larger MAC step size ($n\cdot\mu_0$) should be used. Moreover, attenuating the noise effect to decrease the noise variance (σ'<σ) can also increase the signal margin.

Fig. 3 shows the timing diagram and working principle of the CIM engine. Time-modulated pulses generated by DTC are applied in parallel to the sense lines (SLs) after passing through the pulse path-configuration circuit. Three columns of SRAM cells that store W[2:0] are activated by the MAC input pulses on their SLs. The sign-bit W[3] cell is only used in the sign-control logic to determine which bit line to discharge during the MAC phase, thus the 64 discharging branches embedded in the sign-bit cells can be utilized in the ADC phase. The time-modulated ACT[i]×W[2:0] pulses, whose pulse widths depend on the bit locations of the weight, are applied to SL[2:0] to complete the MAC process. A binary-search readout process is achieved by discharging RBL/B using time-modulated pulses similar to the MAC process. At each step, the SA compares the voltages of RBL and RBLB to determine the digital value of the bit, and the bit-line with a higher voltage is discharged in the next search cycle. The amount of discharged value at each step is shown in Fig. 3 and is controlled by the number of discharging branches activated and the width of the readout-enable pulse. At the end of a 9-bit readout process, RBL and RBLB reach a common voltage value. Since the discharging current $I_0$ is roughly constant, the discharged voltage value on bit-lines is linearly proportional to the accumulated pulse width. Compared with a SAR-ADC of the same bit-precision, the cell-inherent readout mechanism is more energy-efficient because it uses the two bit-line capacitors, which only require to be pre-charged once, for both MAC and readout operations instead of a separate capacitor array that needs full-scale charging or discharging solely for the AD conversion.

Fig. 4 shows the implementation of the proposed two signal margin enhancement techniques. Analysis shows that activation values after ReLU become positive and are concentrated within a range of small values. As the activation value is represented by the DTC output and the simulation result shows that the noise effect is more significant for small pulse width, activations with small value is more susceptible to noise. In the proposed MAC-folding scheme, a constant 8 is subtracted from each ACT value before the MAC operation, and sign-magnitude representation is used. This method reduces the dynamic range of the output at the bit-lines by almost half, which in turn, increases the MAC step size by 1.87×, resulting in an enlarged signal margin. Moreover, since most of the activations are shifted to a range of larger values, the accumulated noise error of analog CIM operation is suppressed. Simulation result using 10 random image inputs shows that the accumulated noise error on the outputs of a convolution layer is 2.51-2.97× smaller. Moreover, statistical results from the simulations indicate that the CIM engine's accumulated MAC results usually do not utilize the entire voltage headroom. In the boosted-clipping scheme, to fully utilize the voltage margin and increase the signal margin, a boosted 2× MAC step is required. This can be achieved by reconfiguring the bias current of DTC to achieve a 2× pulse resolution. With an increased MAC step, there might be values that fall outside the headroom range. Since the MAC result is represented by the voltage difference between RBL and RBLB, a fixed ADC full-scale range ensures only results within the ADC range are quantized. Values larger than the maximum positive or smaller than the minimum negative of the readout range are clipped. The ADC step and full-scale range are configured by combination of DTC pulse width and the number of activated cells for each readout bit.

The 16Kb CIM macro prototype is fabricated using TSMC 40nm process. Fig. 5 shows the measurement settings and the measured performance for different input sparsity. The proposed signal margin enhancement techniques are evaluated by 9K test points of random inputs, the measured 1σ error with 9-bit readout is reduced from 1.3% to 0.64%. The measured transfer curve, differential nonlinearity (DNL), and integral nonlinearity (INL) of the CIM core are also shown. Fig. 6 shows the comparison with state-of-the-art CIM macro designs.

**Acknowledgment:** This research was supported by ACCESS – AI Chip Center for Emerging Smart Systems, sponsored by InnoHK funding, Hong Kong SAR.

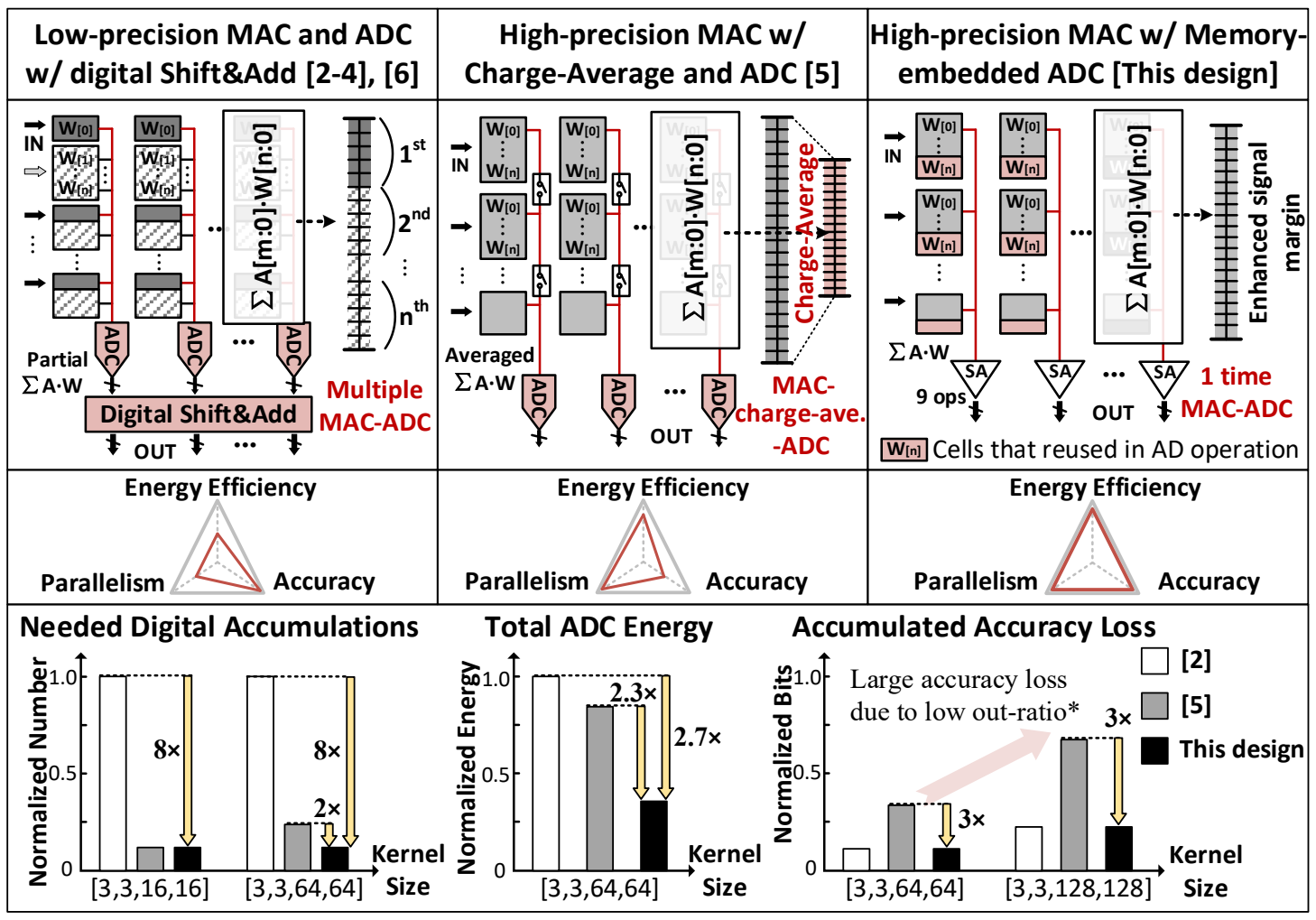

*Comparison is done by mapping a 4-bit ResNet-20 to the CIM cores; SAR-ADC/Readout Energy is obtained by post-simulation with TSMC 40nm; Out-ratio: read-out precision/full output precision[7].

Fig. 1. Comparison with existing CIM designs on parallelism, accuracy, and energy efficiency.

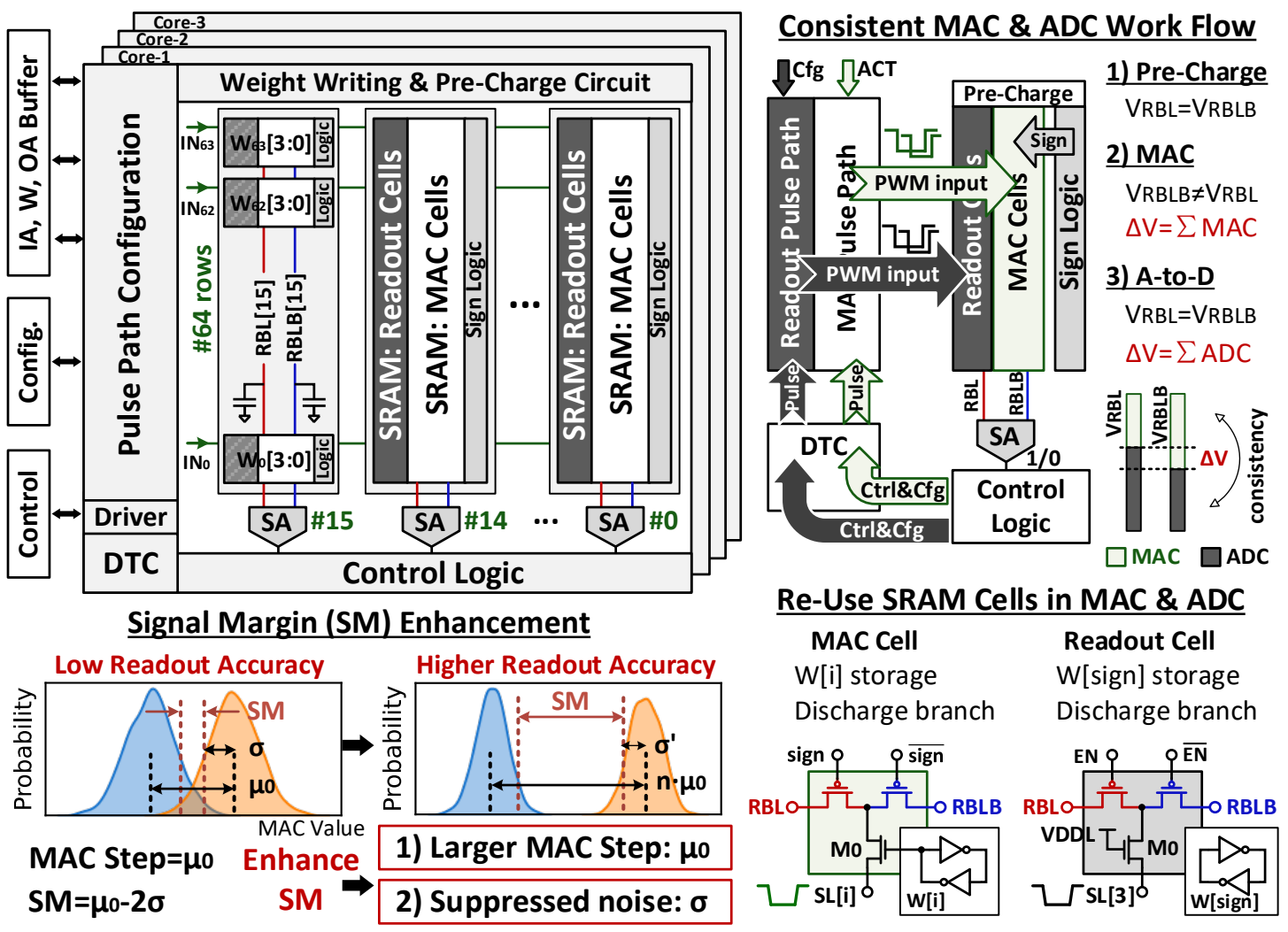

Fig. 2. Overall architecture of the proposed 16Kb CIM architecture with memory cell-inherent MAC and ADC operations; Definition of Signal Margin (SM) and methods for enhancing it.

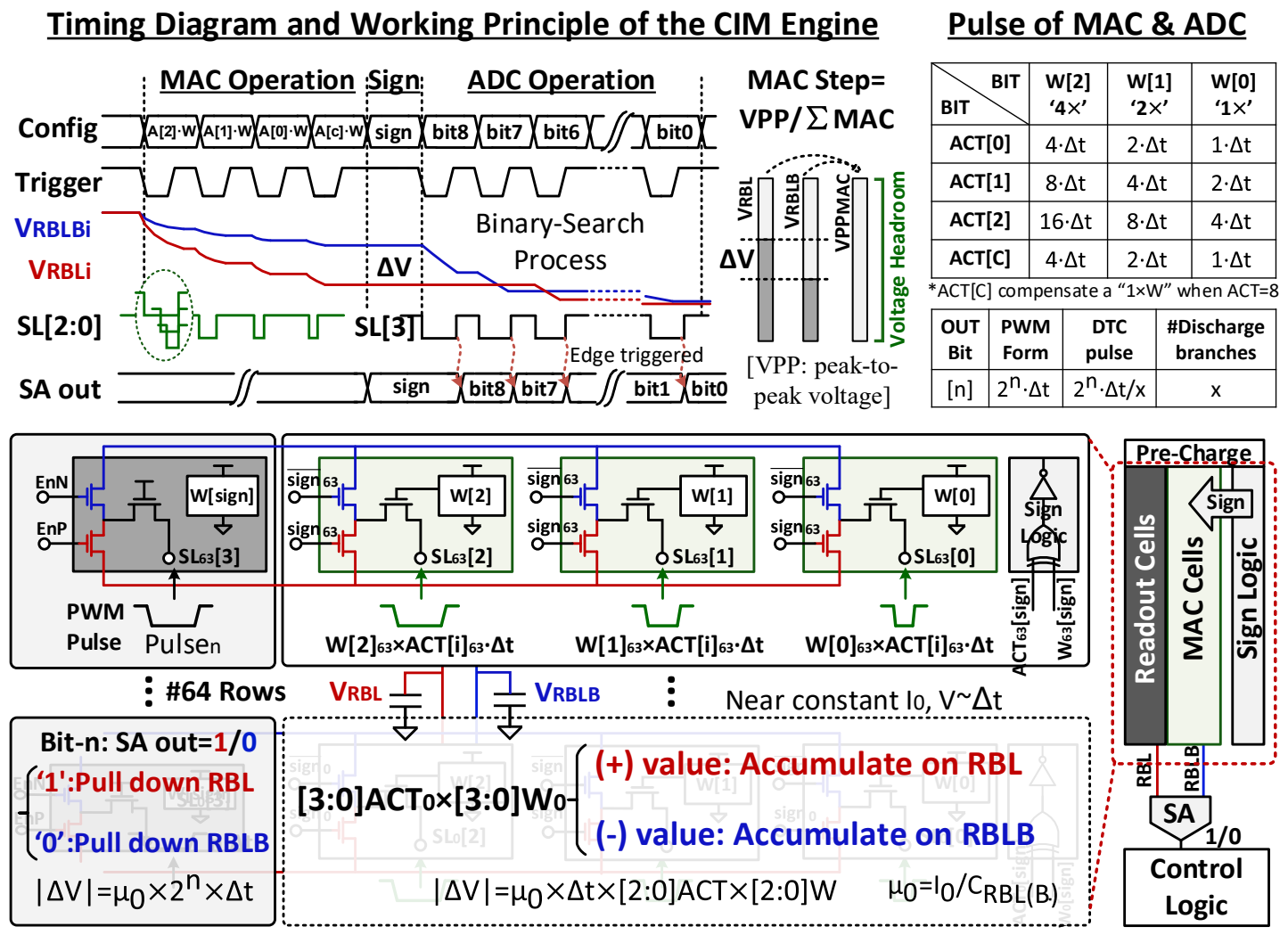

Fig. 3. Timing diagram of the time-modulated MAC and ADC processes; Structure of the column-wise CIM engine and the pulse-integrated MAC and readout principle.

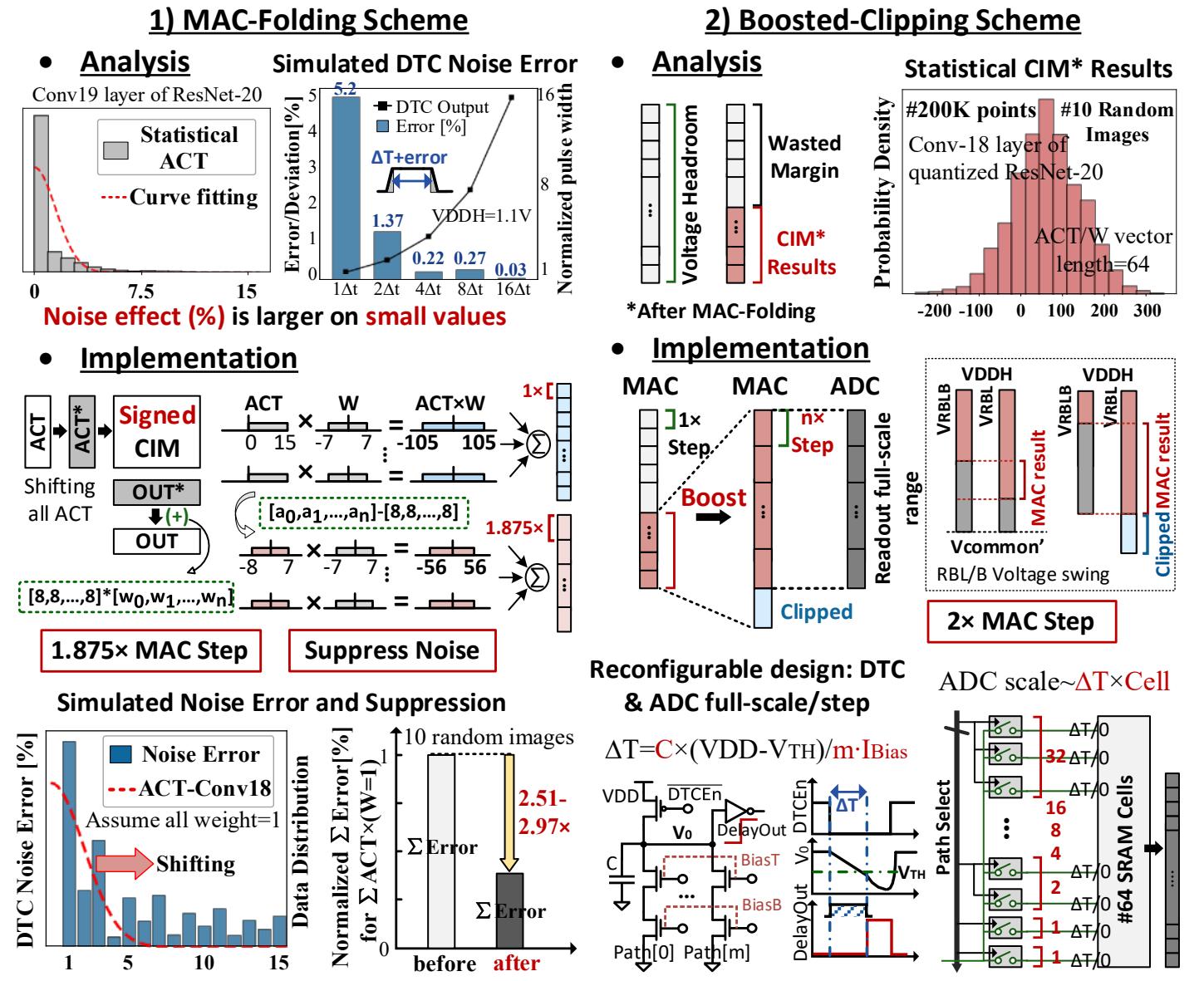

Fig. 4. The proposed two signal margin enhancement techniques.

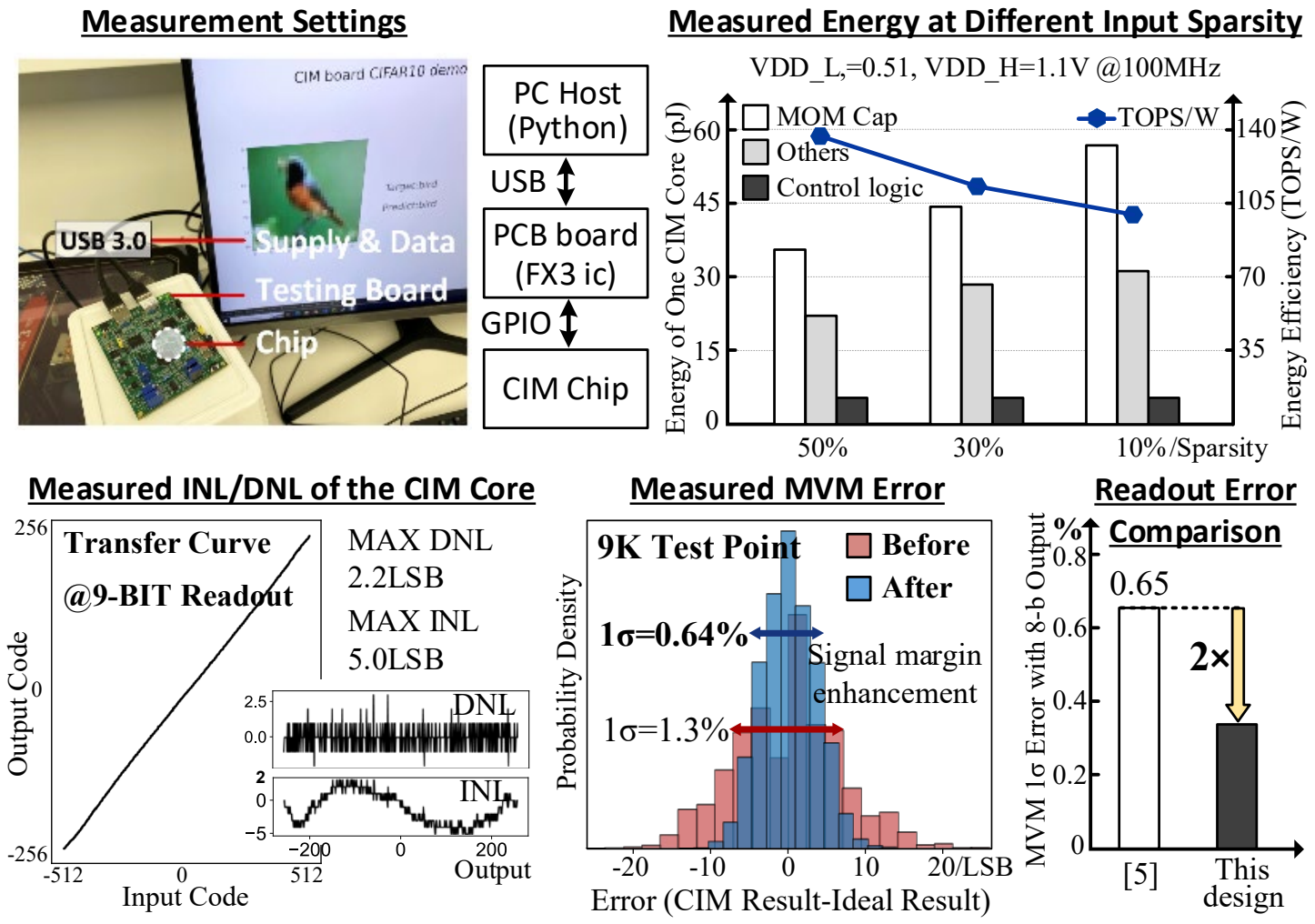

Fig. 5. Measurement settings of the CIM chip; Measured INL/DNL of the CIM engine and performance of the proposed signal margin enhancement techniques.

| | ISSCC'21 [2] | ISSCC'21 [6] | JSSC'22 [3] | VLSI'22 [5] | ISSCC'22 [4] | This Design |
|---|---|---|---|---|---|---|
| Technology (nm) | 28 | 65 | 28 | 22 | 28 | **40** |
| Cell Type | SRAM | SRAM | SRAM | SRAM | SRAM | **SRAM** |
| CIM Memory (Kb) | 384 | 64 | 64 | 128 | 1024 | **16** |
| Clock Freq. (MHz) | - | 25-100 | - | 145-240 | - | **100-200** |
| ACT:W Precision (bit) | [1]4:4 | [1]4:4 | [1]4:4 | 8:8 | [1]4:4 | **[1]4:4** |
| [2]Throughput (GOPS/Kb) | - | 6.17 | - | 4.69-7.81 | 4.15-4.85 | **6.82-8.53** |
| Energy Efficiency (TOPS/W) | 60.28-94.31 | 46.3 | 28-30.4 | 15.5-32.2 | 84.45-112.6 | **95.6-137.5** |
| [3]Area Efficiency (TOPS/W/mm²) | - | 27.1 | - | 62-128.8 | - | **790-1136** |
| 4-b [4]FoM (ACT/W) | - | 4.57 (4b/4b) | - | - | 5.6 (4b/4b) | **10.4 (4b/4b)** |
| 8-b [4]FoM (ACT/W) | - | 1.14 (8b/8b) | - | 1.69 (8b/8b) | 1.39 (8b/8b) | **2.61 (8b/8b)** |

[1]Extendable precision; [2]Normalized with memory size; [3]Normalized energy-based area efficiency [7]; [4]FoM = ACT (bit) × W(bit) × OUT-ratio × Throughput (TOPS/Kb) × Energy Efficiency (TOPS/W). (Average perform.)

Fig. 6. Comparison with state-of-the-arts.

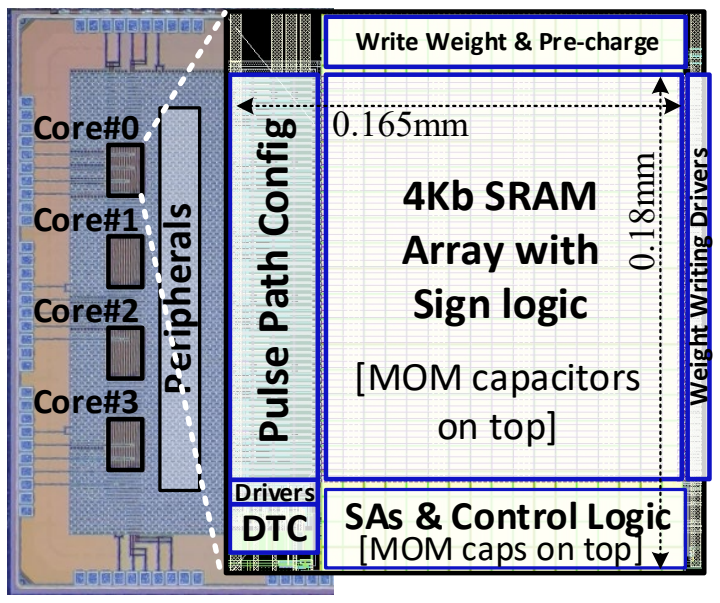

| CHIP SUMMARY | |
|---|---|
| Technology | 40nm CMOS |
| Macro size (Kb) | 16 |
| Core Area (mm$^2$) | 0.121 |
| Analog CIM precision: ACT/W/OUT | 4/4/9 |
| Throughput (GOPS) | 109.2-136.5 |
| [1]Energy Efficiency (TOPS/W) | 95.6-137.5 |
| [2]Inference Accuracy (CIFAR-10) | 91.6% |

[1]Performance with 50%-10% input sparsity
[2]ResNet-20 model, baseline accuracy 92.1%

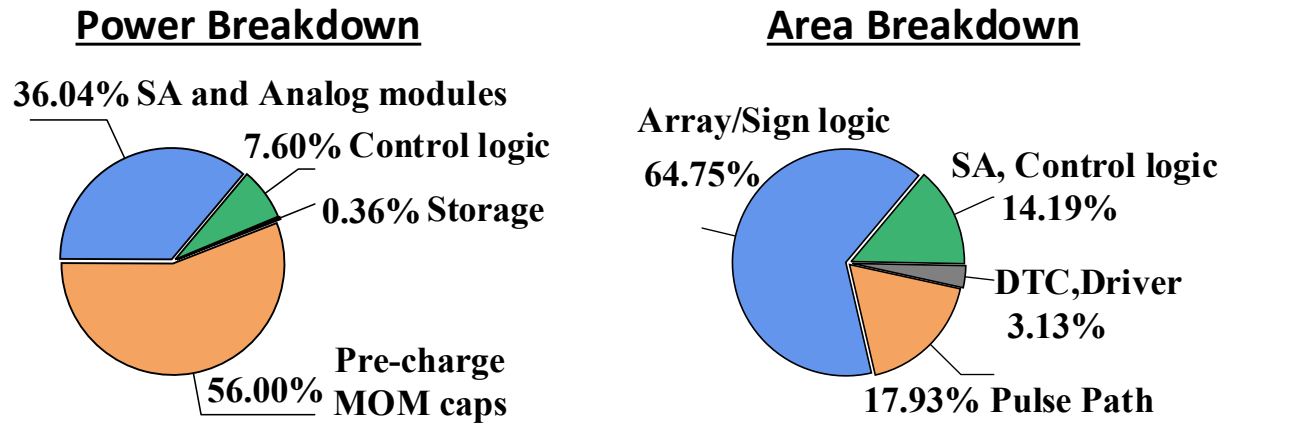

Fig. 7. Die photo, CIM core floorplan and chip summary; Measured power breakdown and the area breakdown.

**Additional References:**